# Full energy spectra of interface state densities for *n*- and *p*-type MoS₂ field-effect transistors

*Nan Fang+, Satoshi Toyoda, Takashi Taniguchi, Kenji Watanabe, and Kosuke Nagashio\**

Dr. N. Fang, S. Toyoda, Prof. K. Nagashio
Department of Materials Engineering, The University of Tokyo 113-8656, Japan
+Present address: RIKEN, Wako, Saitama 351-0198, Japan
E-mail: nagashio@material.t.u-tokyo.ac.jp
Dr. T. Taniguchi, Dr. K. Watanabe
National Institute of Materials Science, Ibaraki 305-0044, Japan

**Abstract:**
Two-dimensional (2D) layered materials are promising for replacing Si to overcome the scaling limit of recent ~5 nm-length metal-oxide-semiconductor field-effect transistors (MOSFETs). However, the insulator/2D channel interface severely degrades the performance of 2D-based MOSFETs, and the origin of the degradation remains largely unexplored. Here, we present the full energy spectra of the interface state densities ($D_{it}$) for both *n*- and *p*- MoS₂ FETs, based on the comprehensive and systematic studies, *i.e.*, thickness range from monolayer to bulk and various gate stack structures including 2D heterostructure with *h*-BN as well as typical high-*k* top-gate structure. For *n*-MoS₂, $D_{it}$ around the mid gap is drastically reduced to $5 \times 10^{11}$ cm⁻²eV⁻¹ for the heterostructure FET with *h*-BN from $5 \times 10^{12}$ cm⁻²eV⁻¹ for the high-*k* top-gate MoS₂ FET. On the other hand, $D_{it}$ remains high, ~$10^{13}$ cm⁻²eV⁻¹, even for the heterostructure FET for *p*-MoS₂. The systematic study elucidates that the strain induced externally through the substrate surface roughness and high-*k* deposition process is the origin for the interface degradation on the conduction band side, while sulfur-vacancy-induced defect-states dominate the interface degradation on the valance band side. The present understanding on the interface properties provides the key to further improving the performance of 2D FETs.

## 1. Introduction

The electric field effect that enables modulation of the carrier density in a semiconductor channel is at the heart of the transistor. The gate controllability represented by the subthreshold swing (*S.S.*), that is, gate-voltage change needed to induce a drain-current change of one order of magnitude, is critical to achieve energy-efficient logic devices.[1] Therefore, for Si metal-oxide-semiconductor field-effect transistors (MOSFETs), many dedicated researchers have developed interface analysis methods based on capacitance-voltage (*C-V*) measurements and have studied the SiO₂/Si interface properties in great detail such that they have become reliable and widely accepted.[2-11] Here, the recent demonstration of a natural thin-body MoS₂ FET with an effective channel length of ~3.9 nm has facilitated research on 2-dimensional (2D) layered channels due to overcoming the scaling limit of ~5 nm for Si gate length.[12] Although the dangling-bond-free surface of the layered channel is expected to ideally provide an electrically inert interface, there are many reports on the wide range of interface state densities ($D_{it}$) from $10^{11}$~$10^{13}$ eV$^{-1}$cm$^{-2}$ for high-$k$ top-gate $n$-MoS$_2$ FET in reality,[13-30] which must be reduced to improve the device performance. To date, several physical origins for $D_{it}$ have been proposed, which are summarized in **Figure 1**a.

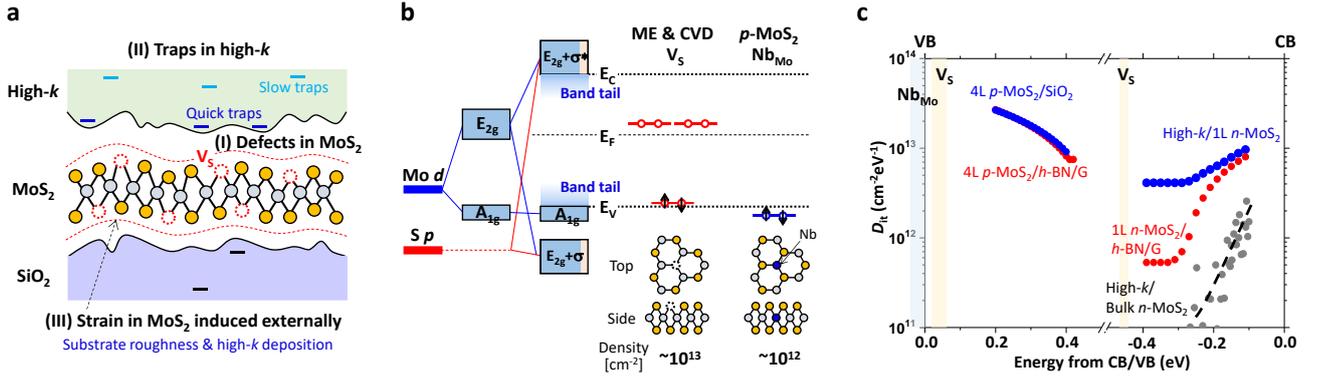

**Figure 1.** (a) Schematic illustration of different origins of interface states in high-$k$/MoS$_2$/oxide structures. (b) Schematic bonding diagram showing how the local orbitals on Mo and S interact to form CB, VB, and interface states in MoS$_2$. V$_S$ and Nb$_{Mo}$ indicate the sulfur vacancy and the substitution of Mo site by Nb, respectively. (c) The full energy spectra of $D_{it}$ for different gate stack structures. Notice that the band gaps of 1L, 4L, and bulk MoS$_2$ are different. Therefore, the transverse axis for the $D_{it}$ - energy distribution is shown as the energy from the CB/VB edge. For the high-$k$/bulk $n$-MoS$_2$ interface, $D_{it}$ is extracted from frequency dispersion-free $C$-$V$ curves by the Terman method.[25] The bulk MoS$_2$ thickness is ~ 58 nm, which is larger than $W_{Dm}$, supporting the availability of the $C$-$V$ measurement. $N_D = 2 \times 10^{17}$ cm$^{-3}$ is used here for a natural $n$-MoS$_2$ crystal, which is precisely determined from thickness-dependent $I$-$V$, $C$-$V$ characterizations.

First, trap (I) represents the defects and impurities in the $n$-MoS$_2$ channel. Sulfur vacancies (V$_S$) with the high density of ~$10^{13}$ cm$^{-2}$ are widely recognized in mechanically exfoliated (ME) and chemically vapor deposited (CVD) MoS$_2$.[31-37] Since physical vapor deposited MoS$_2$ dominantly includes antisite defects,[31] the present study only focus on V$_S$ in ME and CVD MoS$_2$. V$_S$ introduces defect states in the band gap, as shown in **Figure 1**b, which has been evaluated by density functional theory (DFT). CB and VB refer to the conduction band and valance band, respectively. Second, trap (II) represents the traps in the high-$k$ insulator. In general, the back SiO$_2$ oxide is formed by thermal oxidation with well-controlled quality, which usually shows an extremely low trap site density inside (~$10^{10}$ cm$^{-2}$). On the other hand, the top high-$k$ oxide is typically formed on the inert MoS$_2$ surface by atomic layer deposition (ALD) at a relatively low temperature with the aid of a buffer layer, which may introduce many traps inside. The traps close to the interface serve as quick traps while the traps inside the oxide serve as slow traps, as discussed in several reports.[24,28,38] Third, trap (III) represents the strain in MoS$_2$ induced externally. One of the interesting properties of 2D materials is that they can be scaled down to atomic thickness. Strain is easily induced in a thin MoS$_2$ channel by both substrate surface roughness and/or the high-$k$ deposition process, resulting in Mo-S bond bending.[25-27,39] Since the conduction and valence bands of MoS$_2$ are mainly composed of the energy splitting of the Mo $d$ orbital,[40,41] the band tail states will be easily introduced, as schematically illustrated in **Figure 1**b. Although V$_S$ introduces lattice disorder strain around V$_S$ in the ideally flat MoS$_2$ layer, this strain has already been incorporated in the DFT calculation and is regarded as the origin in trap (I). The macroscopic strain introduced externally in the MoS$_2$ layer is considered here in trap (III).

The high-$k$/MoS$_2$ interface properties are inherently complex because $D_{it}$ includes one or more types of traps and some origins might be related to each other. Most of the previous studies only focus on one specific gate stack with limited channel thickness ($t_{ch}$) for $n$-type MoS$_2$. Therefore, a common understanding of the origin for the interface states has not yet been obtained. Although the energy distribution of $D_{it}$ is critical to reveal its origin from the comparison with the DFT calculation, a recent study indicated that the conventional *C-V* method for the $D_{it}$ - energy relation developed for Si systems cannot be simply applied to the FET structure of 2D channels because the channel charging process due to the high channel resistance is more dominant than the electron capture/emission process by the interface trap.[25] In this study, in order to obtain the energy distribution of $D_{it}$, we performed the modeling of $I_D$-$V_{TG}$ characteristics[25,26] by considering the MoS$_2$ channel carrier statistics through the quantum capacitance ($C_Q$) and its transport through the Drude model.[42] Based on the systematic investigation of over 100 devices for both *n*- and *p*- MoS$_2$ with a wide thickness range of 1 layer (L) to bulk and various gate stack structures including a 2D heterostructure with *h*-BN as well as typical high-*k* top gate structure, the whole picture of high-*k*/MoS$_2$ interface is discussed.

## 2. Results and Discussion

### 2.1. Interface States of MoS$_2$ on the Conduction Band Side

A natural *n*-type MoS$_2$ crystal was first studied, whose thin flakes were prepared by mechanical exfoliation. **Figure 2**a shows a schematic drawing of (i) the top-gate MoS$_2$ FET on insulating quartz or SiO$_2$/Si substrates. The quartz substrate with a surface roughness similar to SiO$_2$ was used because the parasitic capacitance can be neglected. The typical drain/source current ($I_{DS}$) – top-gate voltage ($V_{TG}$) characteristics for 1L MoS$_2$ can be found in **Figure 2**c. The two-terminal field-effect mobility ($\mu_{FE}$) are extracted to be 6.2 cm$^2$ V$^{-1}$ s$^{-1}$ under the conditions where the contribution of $C_Q$ is neglected and $C_{ox}$ for the Y$_2$O$_3$ buffer layer (1 nm) and the ALD-Al$_2$O$_3$ (10 nm) is estimated as ~ 0.45 µFcm$^{-2}$ at the accumulation in *C-V*. The hysteresis is relatively small. *S.S.* extracted at the current range of ~10⁻¹² - 10⁻¹⁰ A is mainly discussed in this paper and can be expressed as[1,25]

$$S.S. = \ln 10 \frac{k_B T}{e} \frac{C_{ox} + C_{it}}{C_{ox}},$$

(1)

where $k_B$, $T$, and $e$ are defined as the Boltzmann constant, temperature, and elementary charge, respectively. $C_{ox}$ is the oxide capacitance, and $C_{it}$ is the interface states capacitance ($C_{it} = e^2 D_{it}$). *S.S.* is a simple and effective parameter to directly evaluate $D_{it}$ from 1L to bulk MoS₂ with $t_{ch}$ << maximum depletion width ($W_{Dm}$ = ~48-55 nm).[25] $C_Q$ or depletion capacitance ($C_D$) decreases exponentially with the energy. At the deep subthreshold region, $C_Q(C_D)$ << $C_{it}$, which make them negligible. This relation can be clearly seen in **Figure S4**b (Supporting Information).

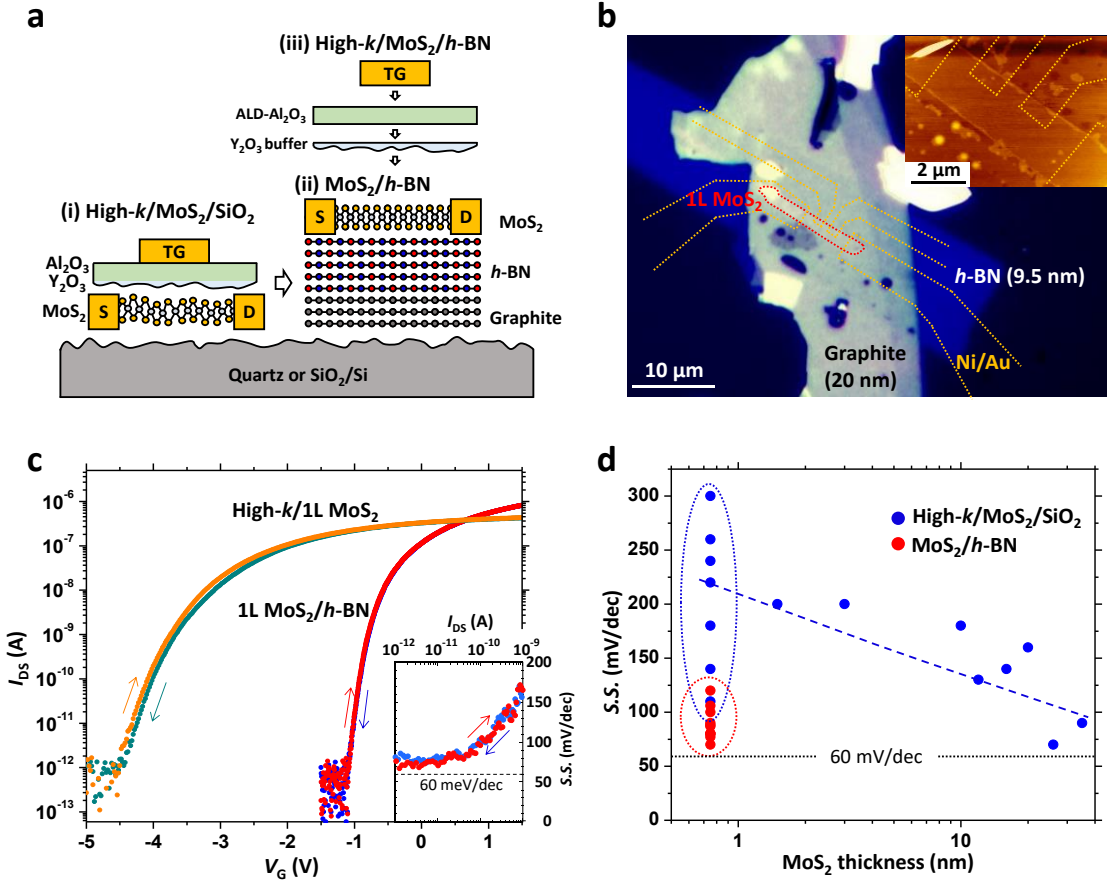

**Figure 2.** (a) Schematics of (i) high-*k*/MoS$_2$/SiO$_2$, (ii) MoS$_2$/*h*-BN, (iii) high-*k*/MoS$_2$/*h*-BN FETs. (b) The optical image of a back-gate MoS$_2$/*h*-BN FET. Inset shows the AFM image. (c) $I_{DS}$–$V_G$ characteristics of high-*k*/1L MoS$_2$ (top-gate sweep), 1L MoS$_2$/*h*-BN FETs (back-gate sweep) at $V_{DS}$ = 0.1 V and RT. Inset shows *S.S.* as a function of $I_{DS}$ for 1L MoS$_2$/*h*-BN FET. (d) *S.S.* as a function of MoS$_2$ thickness for high-*k*/MoS$_2$/SiO$_2$ and MoS$_2$/*h*-BN FETs.

**Figure 2**d summarizes *S.S.* as a function of $t_{ch}$ for a top-gate MoS₂ FET on the quartz substrate. 1L MoS₂ typically has a high *S.S.* level (~230 mV/dec) with a large variation. $D_{it}$ is estimated to be ~8×10¹² cm⁻²eV⁻¹. By increasing $t_{ch}$ over 20 nm, *S.S.* can be reduced as low as ~80 mV/dec with small variation, which corresponds to a $D_{it}$ of ~9×10¹¹ cm⁻²eV⁻¹. The top-gate fabrication process is the same in principle from 1L to in MoS₂, and/or trap (III), the strain in MoS₂ induced externally, are the main origins for the high $D_{it}$ of atomically thin MoS₂. In particular, as for trap (III), it mainly comes from the Mo-S bond bending. Therefore, the thick bulk MoS₂ is robust to this effect, which accounts for the smaller *S.S.* with increasing the MoS₂ thickness.

To clarify the importance of trap (III) and achieve low $D_{it}$ even for the 1L MoS₂ channel toward the ultimately scaled device applications, the MoS₂/*h*-BN/graphite heterostructure FET was prepared as shown schematically by (ii) in **Figure 2**a. The optical image is also shown in **Figure 2**b. The details of the transfer process with negligible bubbles is explained in **Figure S1** (Supporting Information).[43,44] The utilization of back-gate graphite can help with increasing $C_{ox}$ (0.23 μFcm⁻² with dielectric constant of 2.5 for *h*-BN) comparable to that for the high-*k* top gate by reducing the *h*-BN thickness to 9.5 nm, since the atomically flat surface of *h*-BN is guaranteed by the total thickness of graphite and *h*-BN (> 20 nm) (**Figure S1**, Supporting Information). Thus, *S.S.* for the 2D heterostructure FET can be compared with those for the high-*k*/MoS₂ FETs. Moreover, no high-*k* deposition is conducted on the MoS₂ channel to avoid any strain, which bulk, which means that trap (II), traps in the oxide, is unlikely the dominant origin for $D_{it}$ - $t_{ch}$ dependence. Instead, trap (I), the defects indicates that two origins for trap (III), that is, substrate surface roughness and high-*k* deposition, have been suppressed in the present 2D heterostructure FETs. The corresponding $I_{DS}$ – back-gate voltage ($V_{BG}$) characteristics at room temperature (RT) in **Figure 2**c clearly show the improvement of *S.S.* (~75 mV/dec) with negligible hysteresis $\Delta V_{hys}$ = 7.3 mV, where the corresponding $D_{it}$ is ~4×10$^{11}$ cm$^{-2}$eV$^{-1}$. The $\mu_{FE}$ values for two probe (2P) and four probe (4P) measurements are comparable (~70 cm$^2$V$^{-1}$s$^{-1}$) due to an ohmic contact by Ni (**Figure S2**, Supporting Information). More than ten 2D heterostructure FET devices were fabricated and show smaller variation compared with high-*k* top-gate FET devices, as shown in **Figure 2**d. Since 1L MoS$_2$ flakes are mechanically exfoliated from the same natural MoS$_2$ crystals for both high-*k* top gate and 2D heterostructure FETs, trap (I) is unlikely the main origin for the considerably low $D_{it}$. Instead, the low $D_{it}$ is most likely from the suppression of trap (III), which is achieved by forming the atomically flat surface and avoiding the high-*k* deposition process.

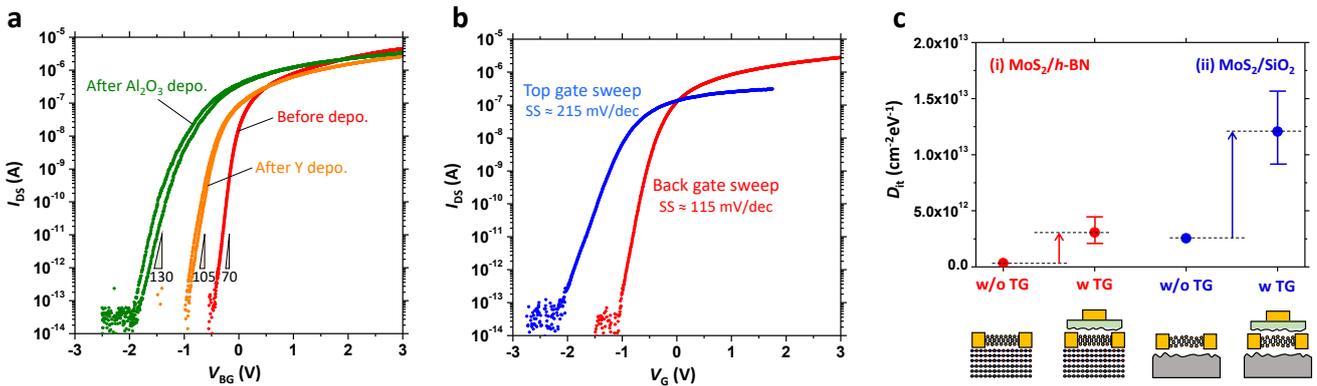

**Figure 3.** (a) $I_{DS}$–$V_{BG}$ characteristics of 1L MoS₂/*h*-BN FET at $V_{DS}$ = 0.1 V and RT before top-gate deposition, after Y₂O₃ buffer layer deposition, and after ALD-Al₂O₃ deposition. (b) $I_{DS}$–$V_G$ characteristics of dual-gate high-*k*/1L MoS₂/*h*-BN FET. (c) The increase in $D_{it}$ by the top-gate deposition process for back-gate 1L MoS₂/SiO₂ FETs and 1L MoS₂/*h*-BN FETs, respectively.

A nearly ideal MoS$_2$/*h*-BN interface is obtained by the 2D heterostructure formation. Here, it is interesting to reveal how the interface is degraded during the high-*k* deposition. Therefore, a 1-nm thick Y$_2$O$_3$ buffer layer, 30 nm-thick ALD-Al$_2$O$_3$ and Al top-gate metal were sequentially deposited on a newly prepared 1L MoS$_2$ heterostructure FET device, as schematically shown by (iii) in **Figure 2**a. The degradation of *S.S.* by the top-gate deposition was monitored through the back interface by the back-gate sweep, as shown in **Figure 3**a. In this device, $C_{ox}$ of *h*-BN with the thickness of 8.1 nm is 0.27 μFcm$^{-2}$ and the thickness of graphite is 29 nm. *S.S.* = ~70 mVdec$^{-1}$ before the top-gate deposition increases to ~105 mVdec$^{-1}$ after the Y buffer layer deposition and finally to ~130 mVdec$^{-1}$ after the subsequent ALD-Al$_2$O$_3$ process. The corresponding $D_{it}$ values are ~3.4×10$^{11}$ cm$^{-2}$eV$^{-1}$, ~1.3×10$^{12}$ cm$^{-2}$eV$^{-1}$, and ~1.9×10$^{12}$ cm$^{-2}$eV$^{-1}$, respectively. The hysteresis also increases to 0.14 V. However, when *S.S.* was measured again before the Al top-gate deposition, *S.S.* unintentionally improved to ~90 mVdec$^{-1}$, as shown in **Figure S3**

(Supporting Information). This result strongly indicates that the degradation of both $D_{it}$ and hysteresis originates mainly from the strain induced by the high-*k* deposition process, not from the defect formation ($V_S$) in MoS$_2$ during the deposition. It should be emphasized that the Y buffer layer is deposited under the Ar pressure of 10$^{-1}$ Pa to entirely suppress the deposition damage in this study.[45,46]

In the case of Si on an insulator (SOI), the top and back interfaces are still separated even for a 7-nm thick Si channel.[47] On the other hand, for a 1L MoS$_2$, high-*k* top-gate deposition considerably affects the back interface. To analyze the top and back interface quantitatively, the top-gate $I_{DS}$-$V_{TG}$ characteristics was measured, as shown in **Figure 3**c. Although the back-gate *S.S.* (115 mVdec$^{-1}$) is better than top gate *S.S.* (215 mVdec$^{-1}$), this result is not fair comparison because $C_{ox}$ for the top and the back are different. Therefore, $C_{ox}$ for a top-gate insulator is extracted to be 0.16 μFcm$^{-2}$ from the top-gate and back-gate capacitive coupling method; then, $D_{it}$ is estimated. It is interesting that the back-interface $D_{it}$ (1.9×10$^{12}$ cm$^{-2}$eV$^{-1}$) is almost the same as the top-interface $D_{it}$ (2.6×10$^{12}$ cm$^{-2}$eV$^{-1}$), which reveals that both top and bottom interfaces must be controlled properly because both interfaces interact with each other for atomically thin 2D channels.

To further separate the contribution of the strain induced by the substrate surface roughness and high-*k* deposition process in trap (III), the degree of the interface degradation by the high-*k* deposition is compared for back-gate 1L MoS$_2$/SiO$_2$/Si and 1L MoS$_2$/*h*-BN/graphite heterostructure FETs. $D_{it}$ is extracted before and after the high-*k* deposition. As shown in **Figure 3**c, $D_{it}$ increases for both substrates after the high-*k* deposition but more drastically degrades for MoS$_2$ on the rough substrate surface, that is, SiO$_2$/Si. The most important finding here is that the degradation of *S.S.* is drastically enhanced in response to the degree of initial substrate surface roughness. Therefore, the strategy to obtain the sharp switching of an atomically thin 2D channel toward the ultimately scaled device applications is to develop stress-free high-$k$ deposition while maintaining the atomic flatness of the substrate surface.

## 2.2. Energy Spectra of Interface States of $n$-MoS$_2$

The interfacial properties discussed above are based on *S.S.* at the deep subthreshold region, which does not provide a precise energy level for $D_{it}$ in the energy band gap. Here, $D_{it}$ is extracted as a function of energy from *I-V* characteristics[25,26] by considering the carrier statistics through $C_Q$ and its transport through the Drude model.[42] An example of this simulation is shown in **Figure S4** (Supporting Information). When the ideal $I_{DS}$-$V_{TG}$ curve is calculated using $C_{ox}$ and $C_Q$ with a constant $\mu$ value extracted experimentally, the ideal *S.S.* of 60 mV/dec can be obtained. The deviation between the ideal and experimental $I_{DS}$-$V_{TG}$ can be considered as the contribution of electron traps to the interface states, that is, $C_{it}$. By assuming the energy distribution of $C_{it}$ as a fitting parameter, the energy distribution of $D_{it}$ of $n$-MoS$_2$ is estimated for high-$k$/1L-MoS$_2$/SiO$_2$ device with top-gate sweep and 1L-MoS$_2$/$h$-BN device with back-gate sweep. On the other hand, for high-$k$/bulk MoS$_2$/SiO$_2$, $D_{it}$ is extracted as a function of energy from *C-V* characteristics by the Terman method.[25] The $D_{it}$-energy distribution is shown in **Figure 1**c. The present analysis can quantitatively capture the band-tail shaped $D_{it}$, which cannot be obtained from the simple $D_{it}$ estimation from *S.S.*. Interestingly, the existence of mid-gap states is evident for high-$k$/1L-MoS$_2$/SiO$_2$ and 1L-MoS$_2$/$h$-BN cases because a constant *S.S.* region exists below the current level of $\sim 10^{-11}$ A, as shown in the inset of **Figure 2**c. Although the mid-gap states are close to the energy for V$_S$, the mid-gap states and band tail states are simultaneously reduced by improving the flatness of the substrate. To further emphasize the substrate flatness instead of Vs effect, **Figure S5**(Supporting Information) shows that $D_{it}$-energy distribution for high-$k$/1L-MoS$_2$/$h$-BN both from top-gate sweep is smaller than that for

high-$k$/1L-MoS$_2$/SiO$_2$ in the whole measured energy range. It is evident that trap (I) is not the main origin for $D_{it}$ in our devices, even though V$_S$ induced defect-states have been widely discussed as the origin for the degradation of the electrical properties for MoS$_2$. As discussed in the previous section, the strain induced by high-$k$ deposition is the dominant origin of the degradation. The interesting point here is that $D_{it}$ near the CB edge for 1L-MoS$_2$/$h$-BN is similar to that for high-$k$/1L-MoS$_2$/SiO$_2$ despite the lack of high-$k$ deposition, which is much larger than that for high-$k$/bulk MoS$_2$. This finding suggests that there is still much room for improvement. At present, the way to improve is not clear yet; the origin may not be the same since the CB minimum changes from the K point for a monolayer to the point on the K → Γ line for the bulk material.[48,49]

## 2.3. Electron Transport of $n$-MoS$_2$

Having clarified the interfacial properties, the transport properties related to the interface states were studied through the temperature dependence for the high-$k$/MoS$_2$/$h$-BN devices. **Figure 4**a shows the temperature dependence of $I_{DS}$-$V_{BG}$ characteristics before the top-gate deposition. The threshold voltage is shifted positively by lowering the temperatures. This is due to the temperature dependence of $C_Q$ of 1L MoS$_2$, as shown in **Figure S4** (Supporting Information), which is further amplified by $D_{it}$. *S.S.* also decreases by lowering the temperature at the range of 150-300 K. This is understood using Equation 1, which is the common case for a bulk Si MOSFET. However, *S.S.* values remain almost constant at 50 -150 K and start to increase at the temperature below 50 K. This result suggests the transition of the electron transport mechanism, that is, the occurrence of interface-states-related transport, such as nearest neighbor hopping (NNH) and variable range hopping (VRH),[34,50,51] because Equation 1 is based on carrier drift/diffusion through the band transport. **Figure 4**b shows $I_{DS}$-$V_{BG}$ characteristics after the top-gate deposition. The threshold voltage shifts positively more substantially by lowering the temperatures, which is due to the increased $D_{it}$. *S.S.* is still proportional to the temperature at the higher temperature region, as shown in **Figure 4**c. In contrast, at the lower temperature region in **Figure 4**b, conductance fluctuation becomes apparent with high repeatability, which further degrades the *S.S.* Since the metal/MoS$_2$ contact is not affected by the top-gate deposition, it cannot be the origin of the observed conductance fluctuation. The conductance fluctuation does not show resonant tunneling behavior.[52] Therefore, this is most likely from the interference of carrier hopping path by localized states.

To further clarify the above scenario quantitatively, **Figure 4**d shows the temperature dependence of $I_{DS}$ at different device. So far, the temperature dependence of $I_{DS}$ for the MoS$_2$ FET is explained by three terms, which are the conduction band transport by thermal activation (TA), and interface-states-related transport by NNH and VRH. Although some papers reported that VRH accounts for the observed temperature dependence of $I_{DS}$,[50,51] we found that NNH gives a reasonable fitting to experimental results at the low temperature, which can be shown by

$$I_{DS} = I_{TA}^0 exp(-E_a/k_BT) + I_{NNH}^0 exp(-E_h/k_BT)$$

, where $I_{TA}^0$, $I_{NNH}^0$ are prefactors. $E_a$ and $E_h$ are the band activation energy and hopping energy, respectively. $E_a$ is the energy difference between the Fermi energy ($E_F$) and the CB minimum, while $E_h$ is the energy

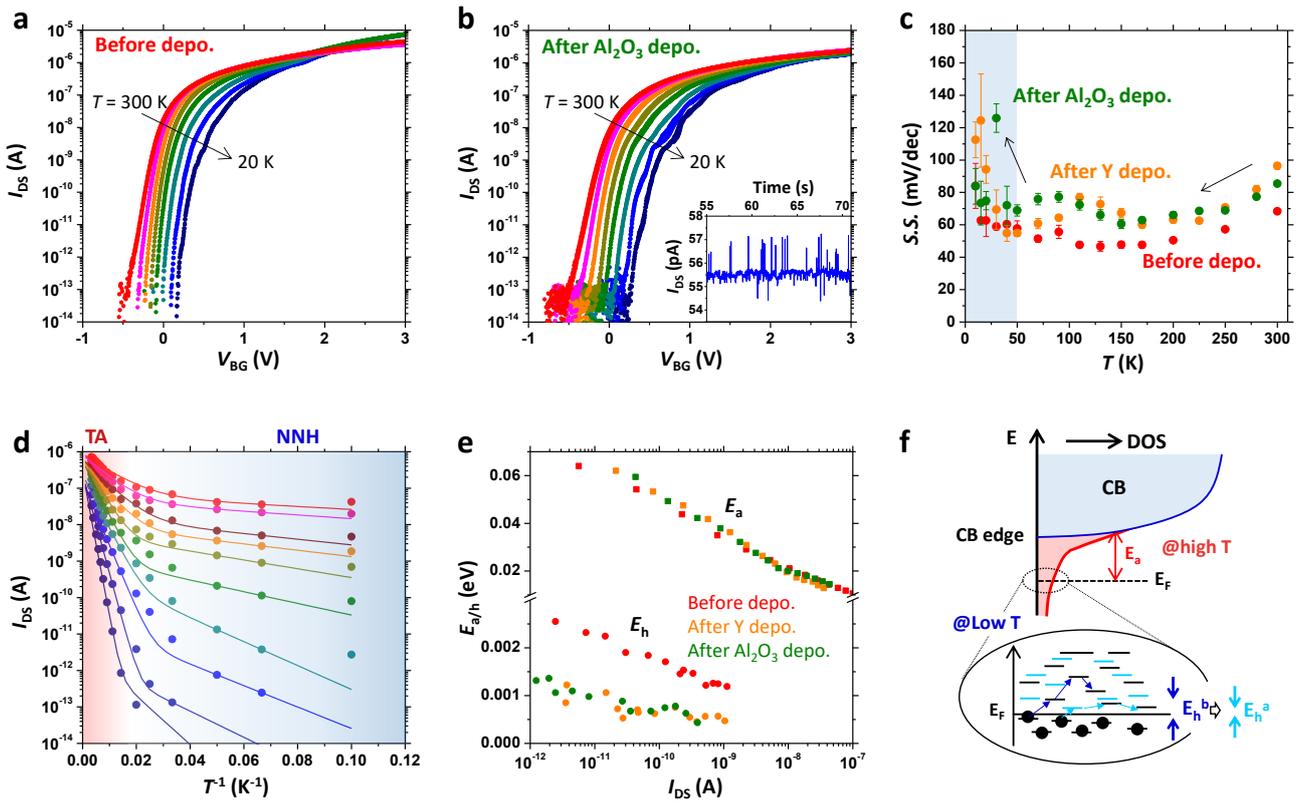

**Figure 4.** $I_{DS}$–$V_{BG}$ characteristics of 1L MoS$_2$/$h$-BN FET at $V_{DS} = 0.1$ V and $T = 300, 250, 200, 150, 110, 70, 40,$ and 20 K (a) before Y$_2$O$_3$ buffer layer deposition and (b) after Al$_2$O$_3$ deposition. Inset shows $I_{DS}$ as a function of time at $V_{DS} = 0.1$ V at 10 K for a different device. (c) *S.S.* as a function of temperature before Y$_2$O$_3$ buffer layer deposition (red), after Y buffer layer deposition (orange), after Al$_2$O$_3$ deposition (green). (d) $I_{DS}$ as a function of $T^{-1}$. The solid circles indicate experimental data, while solid lines indicate the fittings. (e) $E_a$, $E_h$ as a function of $I_{DS}$ before and after the top-gate deposition. (f) Schematic illustration of the energy - *DOS* relation near the CB. The red hatched region just below CB edge indicates the interface states region. Magnified illustration shows the hopping transport through the interface states at low temperature. $E_h^b$ and $E_h^a$ indicate $E_h$ before and after the top-gate deposition, respectively.

$V_{BG}$ values for an Al$_2$O$_3$/Y$_2$O$_3$/MoS$_2$/$h$-BN

difference between the nearest unoccupied interface states. The transition temperature of from TA to NNH (~50 K) is consistent with the transition temperature in the plot of the temperature dependence of *S.S.* (**Figure 4**c). **Figure 4**e shows extracted $E_a$ and $E_h$ values as a function of $I_{DS}$ through the fitting in **Figure 4**d. $E_a$ and $E_h$ are also included for the devices before the top-gate deposition and after the $Y_2O_3$ deposition. $E_a$ remains almost unchanged before and after the top-gate deposition. This supports that the conduction band transport take place at high temperature. In contrast, $E_h$ decreases just after $Y_2O_3$ deposition and no further reduction is observed after $Al_2O_3$ deposition. The decrease in $E_h$ is due to the increase in $D_{it}$, as schematically shown in **Figure 4**f, and accounts for NNH transport at low temperatures at the subthreshold region, that is, the region of conductance fluctuation. The reduced $E_h$ after top-gate deposition could make the hopping path sensitive to Coulombic interactions from localized states. This is then supported by the time-domain of $I_{DS}$ from another device, which shows a similar conductance fluctuation at 10 K, as shown in the insert of **Figure 4**b. Three-level of $I_{DS}$ are dominantly observed at the conductance fluctuation region, which shows multilevel random telegraphic signals (RTSs).[53,54] Since the reduction of $E_h$ for NNH accompanied with the increased $D_{it}$ triggers the observation of RTSs, the initial deposition of insulator, that is, the buffer layer, on the 2D channel must be precisely controlled to avoid introducing strain to the 2D channel.

## 2.4. Interface States of MoS₂ on the Valance Band Side

Natural $MoS_2$ and CVD $MoS_2$ flakes usually show *n*-type behavior as discussed above. Although hole transport has been achieved using ionic gating,[55] the substrate dielectric effect,[56] or a contact metal design,[57,58] these are either thermally unstable or complicated, which makes the investigation of interfacial properties difficult. Therefore, a niobium (Nb)-doped *p*-type $MoS_2$ crystal was studied here, since the substitution of Mo site by Nb ($Nb_{Mo}$) is thermodynamically stable,[59] and uniformly dispersed in $MoS_2$ crystal, which has been confirmed by transmission electron microscopy.[60] **Figure 5**a compares the Raman spectra of *p*-$MoS_2$ and *n*-$MoS_2$. Both types of $MoS_2$ flakes show sharp $E_{2g}$ and $A_{1g}$ peaks and no distinct difference from 1L to the bulk material,[61,62] which indicates that the effect of Nb substitution on the lattice phonon is negligible here.

Back-gate *p*-$MoS_2$ FETs on the $SiO_2/Si$ substrate were then fabricated by following the same procedure as *n*-$MoS_2$, as schematically shown in **Figure 5**b. **Figure 5**c shows $I_D$-$V_{BG}$ characteristics with different $MoS_2$ thicknesses. In contrast to *n*-$MoS_2$, *p*-$MoS_2$ FETs show strong $p^+$-type behavior when $t_{ch}$ > ~8 nm, which is consistent with previous reports,[60] and show ambipolar behavior for 4-nm and 4L-thick devices. Interestingly, the 2L-$MoS_2$ device shows unipolar *n*-type behavior. Two-terminal hole $\mu_{FE}$ and current on-off ratio are 10.5 $cm^2V^{-1}s^{-1}$ and 10 for the 8-nm-thick *p*-$MoS_2$ and 2.5 $cm^2V^{-1}s^{-1}$ and ~$10^3$ for 4-nm-thick *p*-$MoS_2$, respectively. It is consistent with reported hall mobility.[60] Hole $\mu_{FE}$ is further degraded below 1 $cm^2V^{-1}s^{-1}$ for the thinner devices. The off current is dramatically decreased by decreasing $t_{ch}$ because the depletion layer width formed by the gate electrical field becomes close to $t_{ch}$.[25] Therefore, for $t_{ch}$ > $W_{Dm}$, off current remains high, which indicates the existence of unmodulated layers in thick *p*-$MoS_2$ and $W_{Dm}$ can be assumed as ~ 7 nm. Bulk acceptor impurity concentration ($N_A$), which comes from the substitution of the Mo site by Nb, is extracted from $W_{Dm}$ using the relation of

$$W_{Dm} = \sqrt{4\varepsilon_{MoS_2}k_BT ln(N_A/n_i)/e^2 N_A}$$,

where $n_i$ is the intrinsic carrier density with the value of ~ $3\times10^8$ $cm^{-3}$ and $\varepsilon_{MoS2}$ is the dielectric constant of bulk $MoS_2$ in the direction normal to the basal plane, which is 6.3.[63] The extracted $N_A$ is ~ $2\times10^{19}$ $cm^{-3}$, which is consistent with previous Hall measurements.[60] In contrast, the natural *n*-

MoS₂ crystal has bulk donor impurity concentration ($N_D$) of ~ $2×10^{17}$ cm⁻³. [25]

There are two important phenomena observed in Nb-doped MoS₂ FETs; the $p$-type to $n$-type transition and asymmetry in electron and hole transport. The $p$-type to $n$-type transition can be understood by a surface electron accumulation effect. Recently, it was reported that $V_S$ of ~$10^{13}$ cm⁻² is introduced "at the surface" of bulk MoS₂ flake during mechanical exfoliation, and electrons accumulate at the surface.[64] Based on this report, the schematic to explain surface electron accumulation is shown in **Figure 5**d. For the bulk case, bulk $N_A$ is larger than surface $N_D$. By decreasing $t_{ch}$, the $p$-type bulk properties cannot be preserved when the surface $N_D$ becomes larger than bulk $N_A$. To support this idea, $V_{BG}$ values at the charge neutral point ($V_{CNP}$) are plotted as a function of $t_{ch}$ in **Figure 5**e. The charge neutral point (CNP) is defined as the point to change from hole to electron transport in ambipolar behavior (**Figure 5**c). Experimentally, the carrier density at CNP

($V_{CNP}×C_{ox}$) is determined by both the surface donor $N_D^S$ and bulk acceptor $N_A×t_{ch}$. Here, $N_D^S$ (cm⁻²) is assumed as a constant to explain the accumulated electrons at the surface.[64] Therefore, the simple relation of $V_{CNP}C_{ox} = -N_D^S + N_A t_{ch}$ can be obtained.

**Figure 5**e shows $V_{CNP}$ is proportional to $t_{ch}$, which confirms the validity of this relation. Both $N_D^S$ and $N_A$ are calculated to be $8.5×10^{12}$ cm⁻² and $2.7×10^{19}$ cm⁻³, respectively. $N_A$ estimated here is consistent to $N_A$ obtained from $W_{Dm}$. Although $N_A = 2.7×10^{19}$ cm⁻³ is a high doping concentration, it is not high enough to achieve $p$-type transport in 1L MoS₂ because $N_A×0.65$ nm = ~$1.8×10^{12}$ cm⁻² is much smaller than $N_D^S$. As a result, it is the surface doping effect that determines the carrier type at the atomically thin channel.

Another point is the asymmetry between electron and hole transport. **Figure 5**e shows the on-currents of both electrons and holes ($I_{DS}^{ON}$), which are defined as the current at $V_{BG} = +30$ V for electrons and at $V_{BG} = -30$

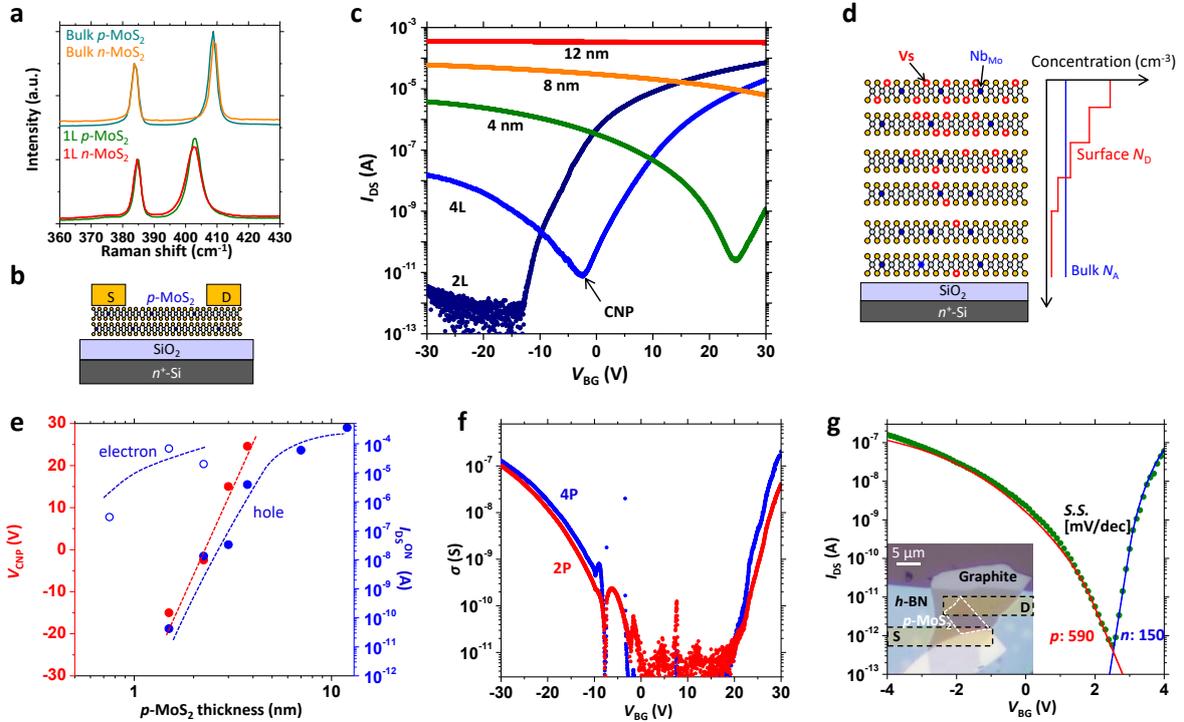

**Figure 5.** (a) Raman spectra of bulk, 1L $n$-MoS₂, and bulk, 1L $p$-MoS₂. (b) A schematic of a back-gate $p$-MoS₂/SiO₂ FET. (c) $I_{DS} - V_{BG}$ characteristics of back-gate $p$-MoS₂/SiO₂ FETs with different $p$-MoS₂ thicknesses at $V_{DS} = 1$ V and RT. (d) Schematic illustration of doping concentrations of bulk $N_A$ and surface $N_D$ as a function of thickness, showing the surface electron accumulation in $p$-MoS₂. (e) $I_{DS}^{ON}$ for both electrons and holes and $V_{CNP}$ as a function of $p$-MoS₂ thickness. (f) $\sigma$ - $V_{BG}$ characteristics of back-gate 4L $p$-MoS₂/SiO₂ FETs for both 2P and 4P measurements at 50 K. (g) $I_{DS} - V_{BG}$ characteristics (circles) and fittings (lines) at $V_{DS} = 1$ V and RT for a 4L $p$-MoS₂/$h$-BN/graphite heterostructure FET with the same structure shown in Figure 2a(ii).

V for holes, respectively. Although the accurate definition should be $I_{DS}^{ON}$ at $V = |V_{BG} - V_{CNP}| = \pm 30$ V, simple definition is used here since $V_{CNP}$ is not seen for thick $p$-$MoS_2$. The $I_{DS}^{ON}$ of electrons is relatively stable with respect to $t_{ch}$, while the $I_{DS}^{ON}$ of holes is degraded dramatically with decreasing $t_{ch}$ even if the $V_{CNP}$ shift is considered. For this $I_{DS}^{ON}$ reduction, the metal/$p$-$MoS_2$ contact effect is carefully investigated since the Schottky barrier at the contact might be the possible origin for the observed degraded hole transport. This is a quite important issue, since $D_{it}$ cannot be simply extracted from $I_{DS}$-$V_G$ if the modulation of the Schottky barrier by $V_G$ dominantly controls the drain current.[65,66] For $n$-$MoS_2$ with a Ni contact, we have already confirmed that the gate controls the high-$k$/$n$-$MoS_2$ interface, not the contact.[25] For $p$-$MoS_2$, the contact effect is discussed as follows.

First, a 4P 4.7 nm-thick $p$-$MoS_2$ FET on a $SiO_2$/Si substrate with Ni contact was fabricated to quantitatively investigate the contact effect. **Figure 5**f shows conductivity ($\sigma$)- $V_{BG}$ characteristics for both the 2P and 4P measurements at 50 K. In general, the contact property becomes dominant at low temperature, since the thermionic-emission current is drastically reduced and the field emission current becomes dominant. In **Figure 5**f, it is clear that the discrepancy between $\sigma_{2P}$ and $\sigma_{4P}$ is quite small. This experiment proves that the contacts for both holes and electrons are reasonably transparent because the doping concentration for both $N_D$ and $N_A$ is high enough. Therefore, for the degraded hole transport observed in **Figure 5**c, the metal/$MoS_2$ contact is not the main origin but an interfacial issue. That is, $D_{it}$ for $p$-$MoS_2$ can be extracted similarly with $n$-$MoS_2$.

To discuss the origin of $D_{it}$ for the VB side according to the classification in **Figure 1**a, 4L-$p$-$MoS_2$/$h$-BN/graphite heterostructure FETs were fabricated for comparison, as shown in the inset of **Figure 5**g. The reason to select 4L is that it is almost the thinnest $MoS_2$ to show $p$-type conduction. Indeed, hole transport can still not be observed for 1L $p$-$MoS_2$ heterostructure with $h$-BN, even though the $S.S.$ for electrons is comparable with the $n$-$MoS_2$ heterostructure. As shown in **Figure 5**g, the $S.S.$ for holes is ~590 mV/dec with $D_{it}$ = ~9.1×10¹² cm⁻²eV⁻¹, while it is ~150 mV/dec with $D_{it}$ = ~1.1×10¹² cm⁻²eV⁻¹ for electrons. Based on the $I_{DS}$-$V_{BG}$ fitting shown by solid lines in **Figure 5**g, the energy spectra of $D_{it}$ for the VB side are summarized in **Figure 1**c, where $D_{it}$ for 4L-$p$-$MoS_2$/$SiO_2$/Si FET is also included. Interestingly, 4L $p$-$MoS_2$ for both $SiO_2$/Si and $h$-BN/graphite substrates show almost equivalent energy spectra for $D_{it}$ with a considerably high level over 10¹³ cm⁻²eV⁻¹ regardless of the substrate flatness. This suggests that both trap (II) and trap (III) are not the main origin because of the lack of high-$k$ deposition and the atomically flat surface of $h$-BN/graphite. Therefore, trap (I), that is, $V_S$ and/or $Nb_{Mo}$, are the most likely cases for $p$-$MoS_2$, because the energy levels for both $V_S$ and $Nb_{Mo}$ are very close to the VB maximum,[58] as shown in **Figure 1**b.

Finally, let us discuss which defect is dominant, $V_S$ or $Nb_{Mo}$. A clue can be observed in **Figure 5**f as the abnormal conductance peaks at the hole side. Very similar abnormal conductance peaks have been reported even in the hole transport for a 1L "$n$-$MoS_2$" FET with ion gating,[67] where it is suggested to be due to defects states by $V_S$ close to VB. **Figure 5**e shows that the degradation of $I_{DS}^{ON}$ for holes as a function of $t_{ch}$ has the same tendency as the $V_{CNP}$ shift. Since the $V_{BG-CNP}$ shift has been suggested to be due to sulfur vacancy formation at the surface, the degradation of $I_{DS}^{ON}$ for holes could also have the same origin, that is, $V_S$. The mid-gap $V_S$ state on the CB side is localized, while the shallow $V_S$ state on the VB side is easily hybridized with valance band states, which will become more prominent in disulfur vacancies and even sulfur vacancy clusters.[68] This explains why the sulfur vacancy-induced transport is more prominent on the VB side (conductance peaks in **Figure 5**f) compared to its transport on the CB side (conductance fluctuation in Figure 4b). Although $Nb_{Mo}$ could also introduce shallow defect states on the VB side in $p$-type $MoS_2$, the density of $V_s$ (~$10^{13}$ cm$^{-2}$)[31] is much larger than that of $Nb_{Mo}$ (~$10^{12}$ cm$^{-2}$) in atomically thin $MoS_2$, which indicates that $V_S$ could be the dominant origin of the degraded hole transport. To improve the hole transport at the subthreshold region, continued efforts to improve the crystallinity are required.

## 3. Conclusion

We successfully extracted the full energy spectra of $D_{it}$ from both $n$- and $p$- $MoS_2$ FETs. By fabricating the 2D heterostructure FET with $h$-BN, on the CB side, it is elucidated that the strain induced externally through the high-$k$ deposition process is the dominant origin of the interface degradation which is further enhanced in response to the degree of initial surface roughness. Therefore, the strategy to obtain the sharp switching for $n$-$MoS_2$ FETs is to develop stress-free high-$k$ deposition while maintaining the atomic flatness for the substrate surface. On the other hand, on the VB side, $V_S$-induced defect states dominate the interface degradation. To improve the hole transport, continued efforts to improve the crystallinity are required.

## 4. Experimental Section

*Device fabrication*: $n$-$MoS_2$ flakes were mechanically exfoliated from natural bulk $MoS_2$ crystals purchased from SPI Supplies, while $p$-$MoS_2$ flakes were obtained from Nb-doped bulk $MoS_2$ crystals purchased from HQ graphene. This Nb doped $MoS_2$ crystals are grown by chemical vapor transport and have been characterized previously.[60] For $MoS_2/SiO_2$ FETs, $MoS_2$ FETs were directly prepared on $SiO_2/Si$ substrates, while for $MoS_2/h$-BN heterostructure FETs, the details of transfer and stacking processes are shown in **Figure S1** (Supporting Information). Ni/Au were deposited as source/drain electrodes for all FETs. For the high-$k$ top-gate formation, 1-nm-thick $Y_2O_3$ was deposited via thermal evaporation of the Y metal in a PBN crucible in an Ar atmosphere with a partial pressure of $10^{-1}$ Pa to form the buffer layer.[45,46] $Al_2O_3$ oxide layers with thicknesses of 10 or 30 nm were deposited via ALD, followed by the Al top-gate electrode formation. No annealing was conducted after the device fabrication. High-$k$/$MoS_2$ FETs were also fabricated on quartz substrate for suppressing the parasitic capacitance in $C$-$V$ measurements.

*Measurements*: Raman spectroscopy and AFM were employed to determine the flake crystal quality and thickness. $I$–$V$ and $C$–$V$ measurements were conducted using Keysight B1500 and 4980A LCR meters, respectively. All electrical measurements were performed in a vacuum prober with a cryogenic system.

## Supporting Information

Supporting Information is available from the Wiley Online Library or from the author.

## Acknowledgements

N. F. was supported by a Grant-in-Aid for JSPS Research Fellows from the JSPS KAKENHI. This research was partly supported by The Canon Foundation, the JSPS Core-to-Core Program, A. Advanced Research Networks, the JSPS A3 Foresight Program, and JSPS KAKENHI Grant Numbers JP16H04343, JP19H00755, and 19K21956, Japan.

**The table of contents entry**

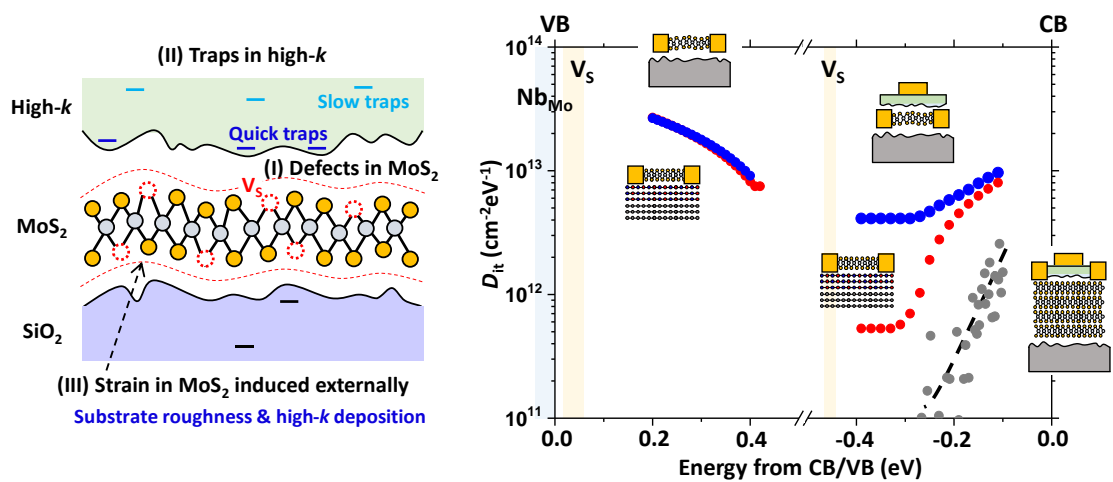

## Supporting Information

# Full energy spectra of interface state densities for *n*- and *p*-type MoS₂ field-effect transistors

*Nan Fang, Satoshi Toyoda, Takashi Taniguchi, Kenji Watanabe, and Kosuke Nagashio\**

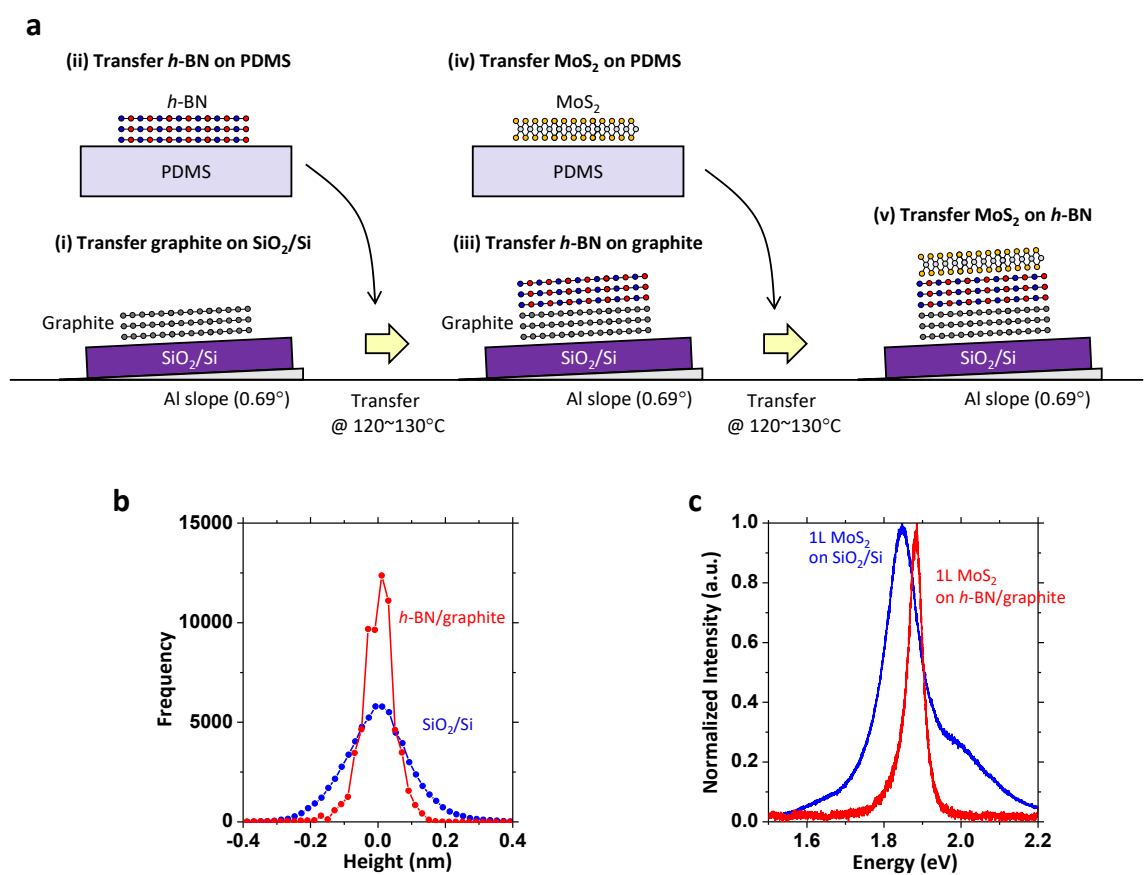

**Figure S1.** (a) Schematics of the fabrication process for MoS₂/*h*-BN/graphite heterostructure. Graphite flakes are first prepared on SiO₂/Si substrate by mechanical exfoliation. *h*-BN flakes are prepared on PDMS by mechanical exfoliation, and then transferred on graphite by using the micromanipulator system with slope of ~ 0.69° at 120 ~ 130°C. MoS₂ flakes are prepared on PDMS by mechanical exfoliation, and again transferred on *h*-BN/graphite to form MoS₂/*h*-BN/graphite heterostructure. (b) Height histogram of *h*-BN/graphite and SiO₂/Si, respectively. (c) Photoluminescence (PL) of 1L MoS₂ on SiO₂/Si and *h*-BN/graphite, respectively. Compared with PL peak on SiO₂/Si substrate, it is sharper on *h*-BN/graphite, which might be attributed to the reduced surface roughness and doping effect.

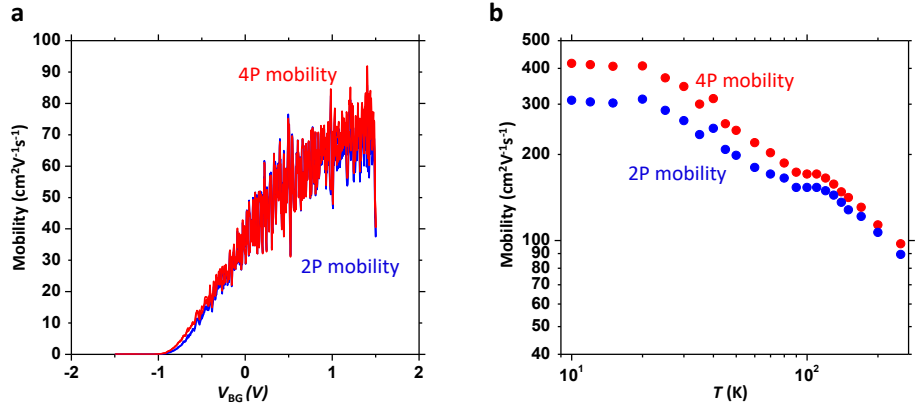

**Figure S2.** (a) 2P and 4P field-effect mobility as a function of $V_{BG}$ for the MoS$_2$/$h$-BN/graphite heterostructure FET. 2P and 4P mobility are comparable with each other, which indicates the ohmic contact by Ni. (b) 2P and 4P field-effect mobility as a function of temperature. A clear phonon-limited temperature dependence is observed, which indicates the high interface properties of the MoS$_2$/$h$-BN/graphite heterostructure.

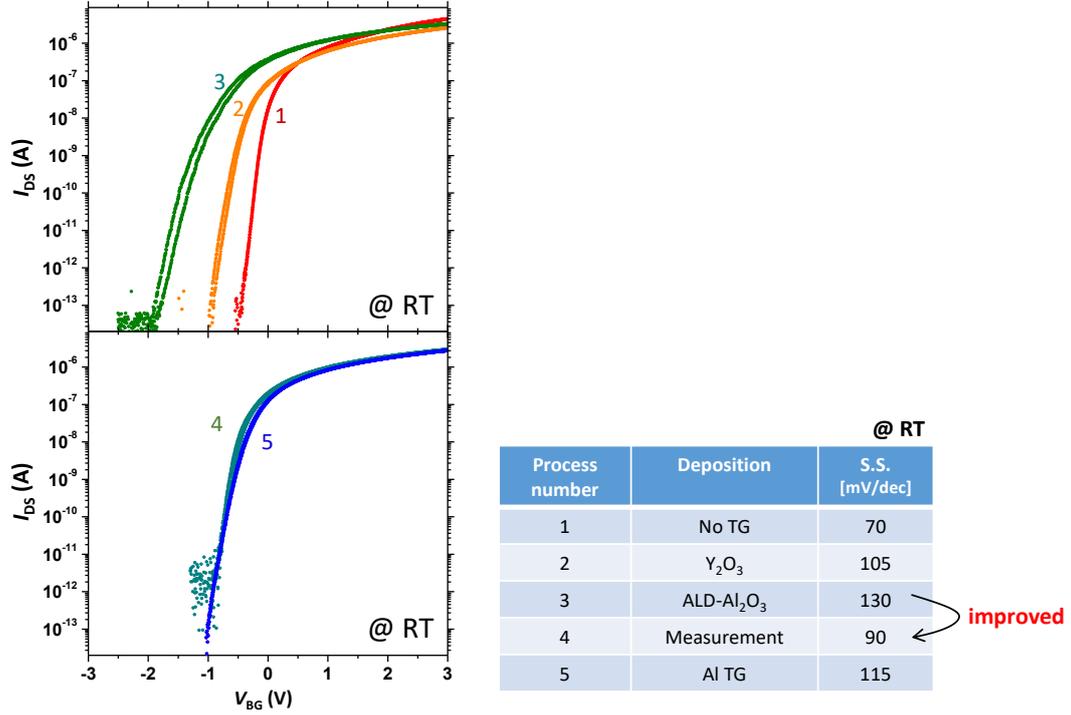

**Figure S3.** $I_{DS}$–$V_{BG}$ characteristics of 1L MoS$_2$/$h$-BN/graphite heterostructure FET at $V_{DS}$ = 0.1 V and RT. The number in the figures indicates the process number in the table. 1: before top-gate deposition, 2: after Y buffer layer deposition, 3: after ALD-Al$_2$O$_3$ deposition, 4: the 2nd electrical measurement, 5: after Al deposition for the top-gate electrode. The sample was always kept in the vacuum desiccator to avoid unintentional doping and so on from the lab environment if the device was not measured just after the top gate fabrication processes.

Interestingly, *S.S.* was unintentionally recovered to ~90 mVdec$^{-1}$ from 130 mVdec$^{-1}$ when this device was measured again after a few days' storage in the vacuum desiccator and before the Al top-gate deposition (step 4). This strongly indicates that the degradation of both $D_{it}$ and hysteresis originates mainly from the strain induced by high-$k$ deposition process, and the strain was released at step 4. Since the maximum temperature during the measurement is RT, the strain might be released by the thermal energy of ~25 meV, which is close to $E_h$ in Figure 4e. Noted that similar improvement of *S.S.* was encountered for other samples.

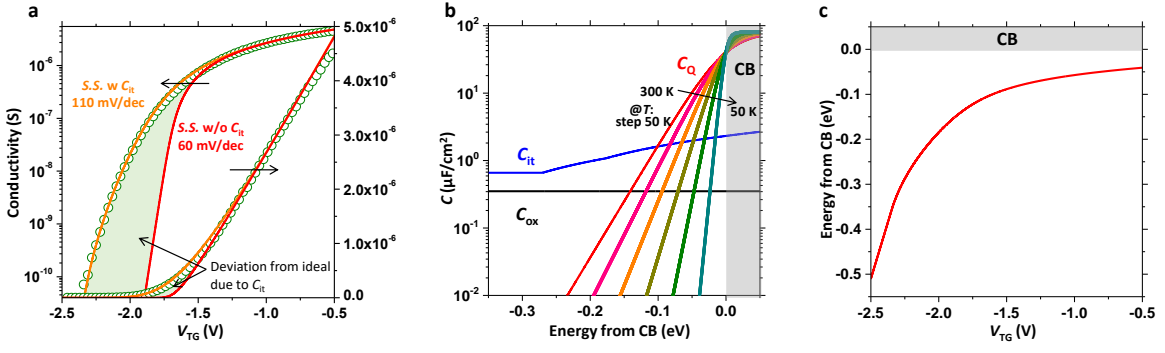

**Figure S4.** (a) $I_{DS}$–$V_{TG}$ characteristics of a high-$k$/1L MoS$_2$/SiO$_2$ FET at $V_{DS}$ = 0.1 V and RT, and the modelling results. (b) The energy distribution of $C_Q$, $C_{ox}$ and $C_{it}$ used in the modelling. (c) $E_F$-$V_{TG}$ relation extracted by the modelling.

**Modelling**: In this model, the electron transport process is considered through the Drude model.[1,2] Conductivity $\sigma = en_{ch}\mu_d$, where $n_{ch}$ and $\mu_d$ refer to channel carrier density and drift mobility, respectively. $\mu_d$ is experimentally extracted as constant field-effect mobility at strong linear region of the $I$-$V$ curve. MoS$_2$ channel carrier density is estimated through oxide capacitance ($C_{ox}$), quantum capacitance ($C_Q$), and interface states capacitance ($C_{it}$). $C_{ox}$ is determined experimentally, while $C_Q$ is theoretically calculated at different temperature using the following relation[1] $C_Q = e^2 g_{2D}\left[1 + \dfrac{\exp(E_G / 2k_BT)}{2\cosh(E_F / k_BT)}\right]$, where $g_{2D} = g_s g_v m^*/2\pi\hbar^2$ is the band-edge *DOS*, and $E_G$ is the bandgap. Here, bandgap $E_G$ = 1.9 eV for monolayer MoS$_2$. $g_s$ and $g_v$ are the spin and valley degeneracy factors, respectively. $m^*$ is assumed to be $0.6m_0$. The mid gap is defined to be $E_F$ = 0 eV. The energy distribution of $C_{ox}$ and $C_Q$ can be seen in (b). When the ideal $I_{DS}$-$V_{TG}$ curve is calculated using $C_{ox}$ and $C_Q$ with constant $\mu$ value extracted experimentally, ideal *S.S.* of 60 mV/dec can be obtained, as shown by solid red line in (a). The deviation between ideal and experimental $I_{DS}$-$V_{TG}$ (hatched region in (a)) can be considered as the contribution of electron traps to the interface states, that is, $C_{it}$. Therefore, $C_{it}$ is used as a fitting parameter to reproduce the $I_{DS}$-$V_{TG}$ curve, as shown by solid orange line in (a). Finally, the energy distribution of $D_{it}$ can be obtained from $C_{it}$ through $C_{it} = e^2 D_{it}$.

It should be noted that extracted $D_{it}$ is still valid even under the constant mobility assumption. The carrier dependent mobility at 300 K has already been calculated in the ref. [1]. Although the mobility increases by more than one order with increasing $n_{ch}$ from $10^{12}$ to $10^{13}$ cm$^{-2}$, it is "almost constant" at the range of $10^{10}$-$10^{11}$ cm$^{-2}$ (subthreshold region). Therefore, present assumption is still valid. Of course, the mobility value itself is much lower than that at the linear region. But, simply say, as can be understood by the fact that $D_{it}$ can be extracted from the *S.S.* value (i.e., the slope of $I_d$-$V_g$), the absolute value of $I_d$ (the mobility value itself) is not important. However, the carrier dependent mobility needs to be considered at linear region.

**Threshold voltage ($V_{th}$) shift**: The nature of $C_Q$ explains many temperature dependent electrical properties of atomically thin MoS$_2$ FET. For example, $V_{th}$ always shift positively with lowering the temperature because $C_Q$ gets close to conduction band (CB). Moreover, this $C_Q$ induced $V_{th}$ shift will be amplified by $C_{it}$, which attributes to the larger threshold voltage shift in Figure 4b compared with that in Figure 4a in the main text.

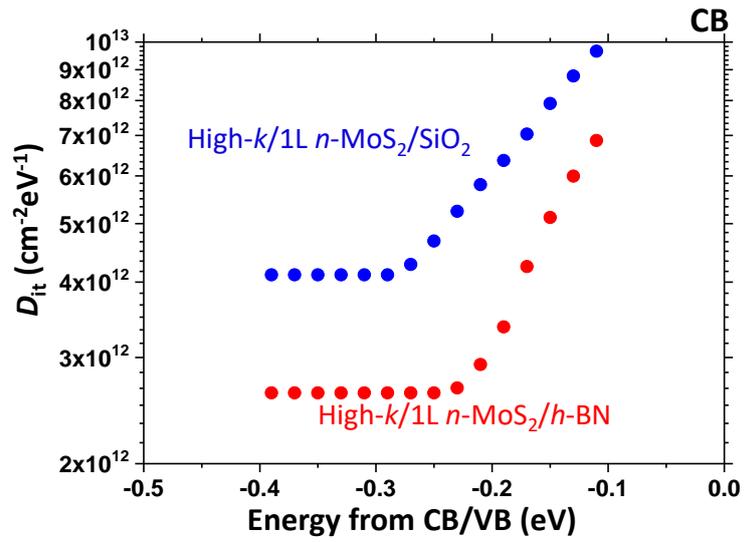

**Figure S5.** The energy spectra of $D_{it}$ for high-$k$/1L $n$-MoS$_2$/SiO$_2$ and high-$k$/1L $n$-MoS$_2$/$h$-BN FETs obtained from top-gate voltage sweep.